\documentclass[aps,prd,twocolumn,twoside,superscriptaddress,floatfix]{revtex4-1}
\pdfoutput=1

\let\oldtabular\tabular
\let\endoldtabular\endtabular
\renewenvironment{tabular}[2][\arraystretch]
  {\edef\arraystretch{#1}
   \oldtabular{#2}}
  {\endoldtabular}

\usepackage{slashed}
\usepackage{amsmath}
\usepackage{multirow}
\usepackage{booktabs}  
\usepackage{graphicx}  
\usepackage{natbib}    
\usepackage{float}

\usepackage{afterpage}
\usepackage{dcolumn}
\usepackage{bm}
\usepackage{amssymb,amsmath}  
\usepackage{lipsum}
\usepackage{setspace}
\usepackage{etoolbox}

\newcommand{\apjl}{ApJL}

\newcommand{\aap}{A\&A}
\newcommand{\jcap}{JCAP}
\newcommand{\mnras}{MNRAS}

\newcommand{\be}{\begin{equation}}
\newcommand{\ee}{\end{equation}}

\begin{document}
\title{Measuring the Duration of Last Scattering}
\author{Boryana Hadzhiyska}
\affiliation{Department of Astrophysical Sciences, Princeton University, Princeton, NJ 08540, USA}
\author{David N. Spergel}
\affiliation{Department of Astrophysical Sciences, Princeton University, Princeton, NJ 08540, USA}
\affiliation{Center for Computational Astrophysics, Flatiron Institute, New York, NY 10010, USA}

\begin{abstract}
The cosmic microwave background (CMB) fluctuations effectively measure the 
basic properties
of the universe during the recombination epoch.  CMB measurements
 fix the distance to the surface of last scatter,
the sound horizon of the baryon-photon fluid and the fraction  of the 
energy density in relativistic species.  We show that
the microwave background observations can also very effectively constrain the
 thickness of the last scattering surface, which is directly related to the
 ratio of the small-scale E-mode polarization
 signal to the small-scale temperature signal.
The current cosmological data enables a 0.1\% measurement
 of the thickness of the surface of last scatter: $19 \pm 0.065$ Mpc.
This constraint is relatively model-independent, so it can provide 
a new metric for systematic errors and an
 independent test of the $\Lambda {\rm CDM}$ model.
On the other hand, it is sensitive to models which affect the
reionization history of the universe such as models with annihilating 
dark matter and varying fundamental constants (e.g., the fine-structure
constant, $\alpha_{\rm EM}$, and electron rest mass, $m_{\rm e}$)
 and as such can
be used as a viable tool to constrain them.
\end{abstract}

\maketitle 
\section{Introduction}
The cosmic microwave background (CMB) stores a tremendous amount of 
information about
the early history of the Universe and its subsequent evolution. Its 
large-scale anisotropies indicate
the existence of tiny fluctuations in the primordial gravitational 
potential
that are the seeds for the formation of galaxies and other large-scale structure. 
Shortly after the discovery of the CMB in 1965, it was also shown
 \cite{1968ApJ...153L...1R} 
 that anisotropic Thomson scattering of photons and electrons 
induces a degree of linear polarization in the data.
The polarization properties of the CMB provide
yet another set of observables, the measurement of which enriches
greatly our understanding of
the Universe.

The CMB
anisotropies were formed primarily around the epoch of hydrogen
recombination 
around redshift $z \sim 1100$ 
defined by the peak of the visibility function,
 $v = |\dot \tau| e^{-\tau}$, which
describes the probability that a CMB
photon last scattered off free electrons
at a particular point in the history of the Universe. 
The shape of the CMB power spectrum is thus most
sensitive to changes around the peak of the visibility
function.
For example, the location of that peak
 determines the distance to the last scattering
surface, which in turn determines the positions of the
peaks of the CMB power spectra.
If we were to increase the width of the visibility
function, that would correspond to a prolonged period of recombination,
 leading to more Thomson scatterings
of photons off free electrons. On scales smaller than the
recombination width, these scatterings lead to
the cancellation of the CMB anisotropies along the line of
sight, while on larger scales they lead to enhancement of
 the polarization signal.  Similarly, changes
in the ionization history, the photon-baryon sound speed, 
the gravitational potential around matter-radiation equality and other
primordial properties which we have no
direct way of probing
would likely lead to measurable changes in the CMB
power spectra and the baryon acoustic oscillation (BAO) peak,
 which would then allow us to put constraints on the features of the 
early Universe \cite{harari}. For example, models which involve
 annihilation of dark matter
to Standard Model particles between the period of 
recombination and reionization result in a modification of
the ionization history, as they lead to a
heating of the baryons and ionization of the neutral
hydrogen, and thus alterations in the CMB visibility function.
Another such example are models which predict variations 
of the fundamental constants --
e.g., the fine-structure 
constant, $\alpha_{\rm EM}$,
and the  electron rest mass, $m_{\rm e}$ provide another 
 such  example and  can thus directly impact CMB observables
\cite{1999PhRvD..60b3516K,2000PhRvD..62l3508A,2001PhRvD..63d3505B,
2001PhRvD..64f3505A,2004MNRAS.352...20R,2009arXiv0906.0329S,
2010HiA....15..307S,2012PhRvD..85j7301M}. Due to their nature of affecting mostly the large-scale
observables, constraints on such models are not expected to
 become much more stringent with the drastic improvement in
 sensitivity expected of future cosmological 
surveys.

Over the past decade, cosmologists have been exploring many alternative ways
to test our understanding of the early Universe. A validated approach they 
have taken
is to introduce new parameters into the analysis of cosmological data,
which can serve as powerful probes for discrepancies with our predictions
and can help in the detection of systematic errors 
in our measurements or
problems with the $\Lambda {\rm CDM}$ model.
In principle, to show that a given new parameter may be of such use, 
one needs to check
that it is not correlated with the standard parameters, i.e. that it reflects
a different physical effect.
The degree of correlation between the new and the standard parameters can 
be tested
by studying their contour plots in a Monte-Carlo engine analysis 
\cite{2013ascl.soft07002A}.

An example of a parameter which probes the properties of the early Universe
 is the width of the visibility function
during the ``last scattering'' of photons,
By changing its width
the strip of time from which the photons could have come would be broadened 
or narrowed. 
A longer period of last scattering would then result in a more polarized 
signal on large scales,
since the photons would have scattered off the electrons
 a larger number of times \cite{harari,hubyharari}.
Therefore, an intriguing question to explore is: how well can we constrain
the width of the last scattering surface with the most recent polarization
 data from the $Planck$ team.
So far, the width of the last scattering surface has not been measured 
directly by CMB experiments,
but its theoretical value is readily computed to be 
$\Delta \eta_\ast \approx 19 \ \rm{Mpc}$ ($\Delta z_\ast \approx 90$)
 using the latest cosmological codes, e.g. CLASS \cite{2011JCAP...07..034B},
 and  standard values for the cosmological parameters from 
\textit{Planck} \cite{Ade:2015xua}.

This paper is organized as follows. We first provide motivation for our work
by stating a relationship between the polarization-to-temperature
ratio of the
power spectra and the width of the visibility function pointed out by
 \cite{harari}.
We then parametrize this width through $\alpha_{\rm vis}$ and explore
how varying this new parameter affects the observable power spectra. 
Finally, we put constraints on its value using the latest CMB data from 
$Planck$, study its degeneracy
with other parameters,
and discuss its potential as a model-independent test of the $\Lambda \rm{CDM}$
 model and of systematic errors. It also provides a powerful tool to constrain models
which alter the reionization history of the Universe such as dark matter annihilation
and decay and variable $\alpha_{\rm EM}$ and $m_{\rm e}$  models.

\section{Physical Motivation}
\label{sec:width}
The shapes of the temperature and polarization power spectra are 
affected by the width of
the last scattering surface  due to \textbf{diffusion (Silk) damping}
 \cite{1968ApJ...151..459S}. Diffusion damping results from
the scattering of photons off electrons during the free-streaming
 epoch \cite{hubyharari}.
The collisions of the free-streaming dipole 
produce a quadrupole moment in the photon distribution function,
 which in turn leads to a polarization
 of the CMB, as the polarization is proportional to the quadrupole
 moment of the photon distribution function
 \cite{harari,hubyharari}.
The polarization of the CMB is produced during the
process of decoupling of matter and radiation, and is also 
proportional to the width of the last
scattering surface $\Delta \eta_\ast$ and the conformal time of 
recombination $\eta_\ast$.
The anisotropy in the CMB temperature also depends on these 
quantities, but
differently. For instance, on large  scales it is very insensitive
  to the values of
$\Delta \eta_\ast$ and $\eta_\ast$  \cite{harari}.

The ratio of the polarization spectrum to the temperature 
spectrum is shown to be
strongly dependent on the width of the last scattering
 surface \cite{harari}:
\begin{equation}
\frac{C_\ell^{EE}}{C_\ell^{TT}} = \Big[ 
\frac{\Delta \eta_\ast \eta_\ast}{(1+R)r^2} \Big]^2  A_\ell,
\label{eq:rat}
\end{equation}
where $r$ is the distance to the last scattering surface, 
$A_\ell$ is a scale-dependent factor, 
and $ R= 3 \rho_b / 4 \rho_\gamma$ is
the ratio between baryonic and radiation densities.

\section{Methods and Tools}
\subsection{Parametrization}
\label{sec:avis}
We parametrize the width of the visibility function,
 $\Delta \eta_\ast$, with our
new parameter $\alpha_{\rm vis}$ in the following way.

\begin{figure}
[H]
      \includegraphics[width=3.59in]{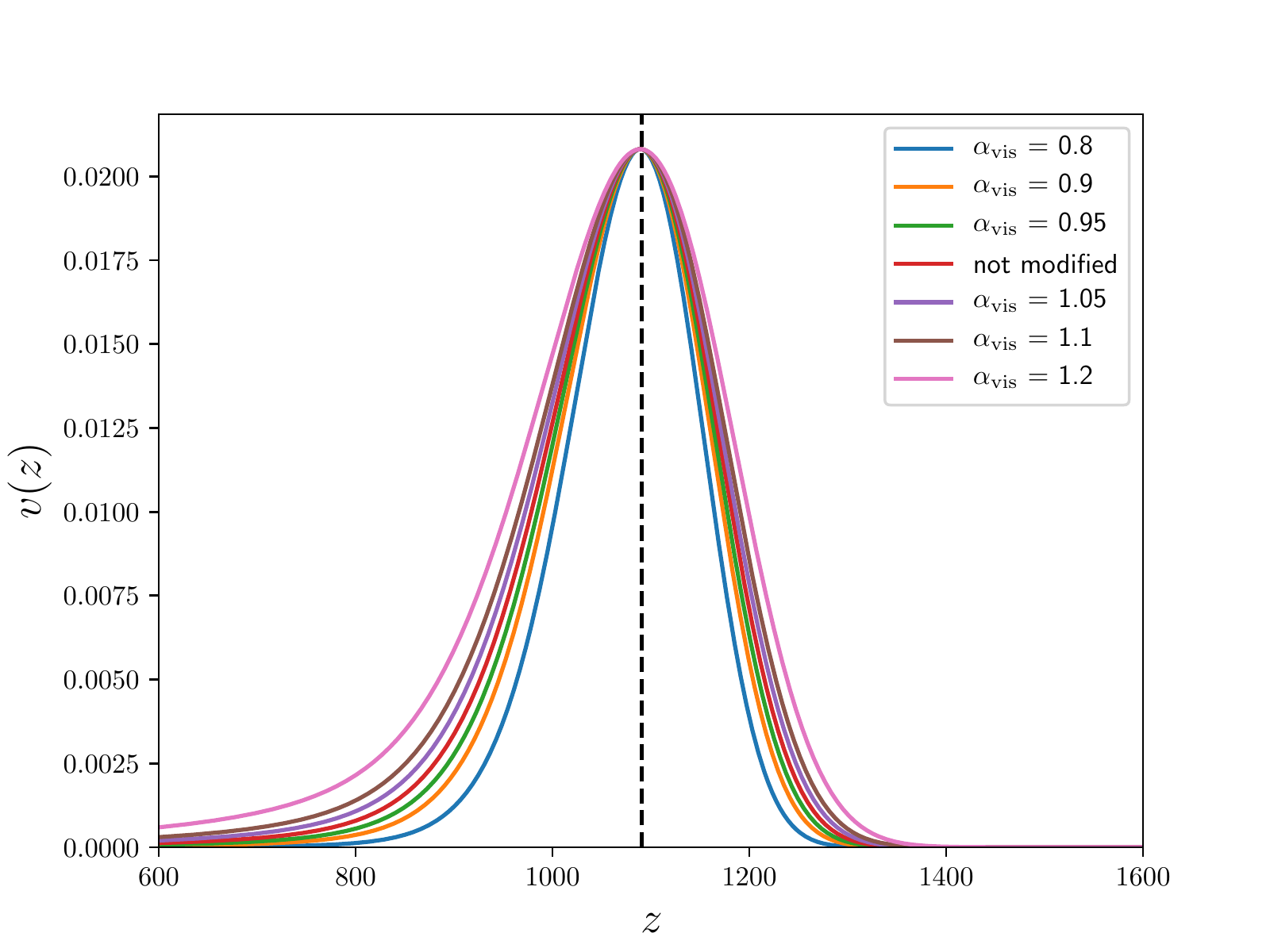}
      \caption{The effect of changing the parameter $\alpha_{\rm vis}$ on 
        the numerically derived visibility function from
	CLASS \cite{2011JCAP...07..034B} 
	as a function of redshift. This parametrization preserves very 
        well the shape of the visibility function and only changes its
        width. CLASS automatically normalizes the area under the
 	visibility function to 1.}
      \label{fig:alpha}
\end{figure}

The visibility function $v_{\rm orig}$
 is well approximated by a Gaussian of width
 $\sigma \equiv \Delta \eta_\ast$ with some
maximum value at recombination $v_{\rm max}$ \cite{1985A&A...149..144J}:
\begin{equation}
v_{\rm orig} \approx v_{\rm max} \frac{1}{\sqrt{2 \pi \sigma^2}} e^{-\frac{ (\eta-\eta_\ast)^2}{2 \sigma^2}} .
\end{equation}
To change the width of a Gaussian, simply multiply its width by
a constant, i.e. $\sigma \rightarrow \alpha_{\rm vis} \times \sigma$.
In this case, the new visibility function becomes:
\begin{equation}
v_{\rm new} \approx v_{\rm max} \frac{1}{\sqrt{2 \pi (\alpha_{\rm vis} \sigma)^2}} 
e^{-\frac{(\eta-\eta_\ast)^2}{2 (\alpha_{\rm vis} \sigma)^2}} .
\end{equation}
However, notice that we can write this as:
\begin{equation}
v_{\rm new} = \frac{v_{\rm max}}{\alpha_{\rm vis}} \left( \frac{v_{\rm orig}}{v_{\rm max}} 
\right)^{1 \over \alpha_{\rm vis}^2}  .
\end{equation}
which can be applied to all Gaussian-like functions,
 as it does not assume anything specific about the function
apart from an approximately Gaussian shape \footnote{We thank Alwin Mao
 for coming up with the idea of this parametrization.}.
The results of this parametrization are shown in Fig. \ref{fig:alpha}.

\subsection{Data Analysis}
We constrained the width of the last scattering surface, $\alpha_{\rm vis}$,
 utilizing the Monte-Carlo sampling engine MontePython 
\cite{2013ascl.soft07002A}
with the \textit{Planck} 2015 measurements of the CMB power 
spectra \cite{Adam:2015rua}
through two sets of likelihoods:

{\bf The first set} constrains it through the Silk damping effect and 
the large-scale polarization data, using only the {\it lite} high $\ell$ likelihoods:  
\begin{itemize}
\item{\bf low $\ell$} -- consists of the CMB TT, EE, BB and TE
 power spectra from $\ell=2$ to 29 (inclusive), and an extra nuisance parameter
 for the overall Planck calibration.
\item {\bf high $\ell$ lite} -- consists of the CMB TT power spectrum from 
$\ell=30$ to 2508 and 
the Planck absolute calibration nuisance parameter.
\item {\bf lensing} -- consists of the $\phi \phi$ lensing spectrum for $\ell=1$
 to 2048 (inclusive)
\end{itemize}

{\bf The second set} uses both the full temperature and E-mode polarization high
 $\ell$ spectra
 in order to constrain the width of the visibility function
 from the ratio of polarization to temperature: 
\begin{itemize}
\item{\bf low $\ell$}
\item{\bf high $\ell$ TTTEEE} -- consists of the CMB TT, EE and TE power spectra
 from $\ell=30$ to 2508 (inclusive), 
and a vector of 94 nuisance parameters.
\item{\bf lensing}
\end{itemize}

A simple $\chi^2$ test shows that current data should constrain $\alpha_{\rm vis}$ to 
less than 1\% of the 
width of the visibility function, 
 so we select a uniform prior on $\alpha_{\rm vis} \in (0.97, 1.03)$. The duration of the 
last scattering surface is
then quantified by the product $\alpha_{\rm vis} \Delta \eta_\ast$, where $\alpha_{\rm vis}=1$ 
corresponds to the Standard
Model prediction for its duration. 

\section{Results}
\label{chap:resls}
\subsection{Width of the Last Scattering Surface}
\begin{figure}[H]
      \includegraphics[width=3.5in]{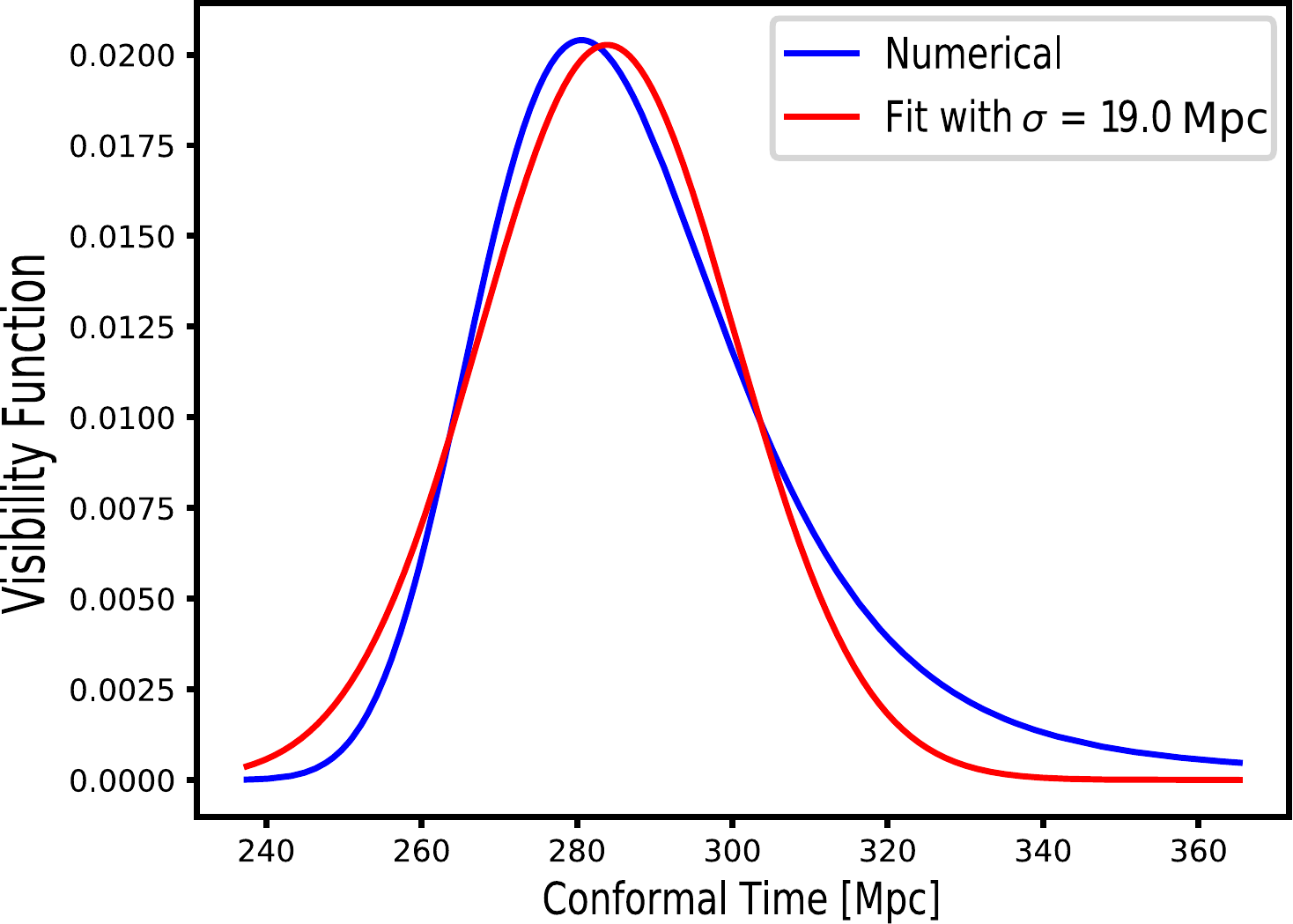}
      \caption{Gaussian fit for the visibility function. The width of the visibility
	function seems to be well-approximated by
      $\Delta \eta_\ast \approx 19 \ {\rm Mpc}$.}
      \label{fig:Gaussian_fit}
\end{figure}

In Fig. \ref{fig:Gaussian_fit}, we show the visibility function $v = |\dot \tau| e^{-\tau}$
 as output by CLASS \cite{2011JCAP...07..034B} and the
best-fitting Gaussian function
with mean $\eta_\ast$ and width $\Delta \eta_\ast$. We 
find the width of the Gaussian fit to be approximately $\Delta \eta_\ast \approx$
19 Mpc (or $\Delta z_\ast \approx 90$ in redshift space), where we used
 standard values for the cosmological parameters \cite{Ade:2015xua}.

\begin{figure*}
      \centering
      \includegraphics[width=7in]{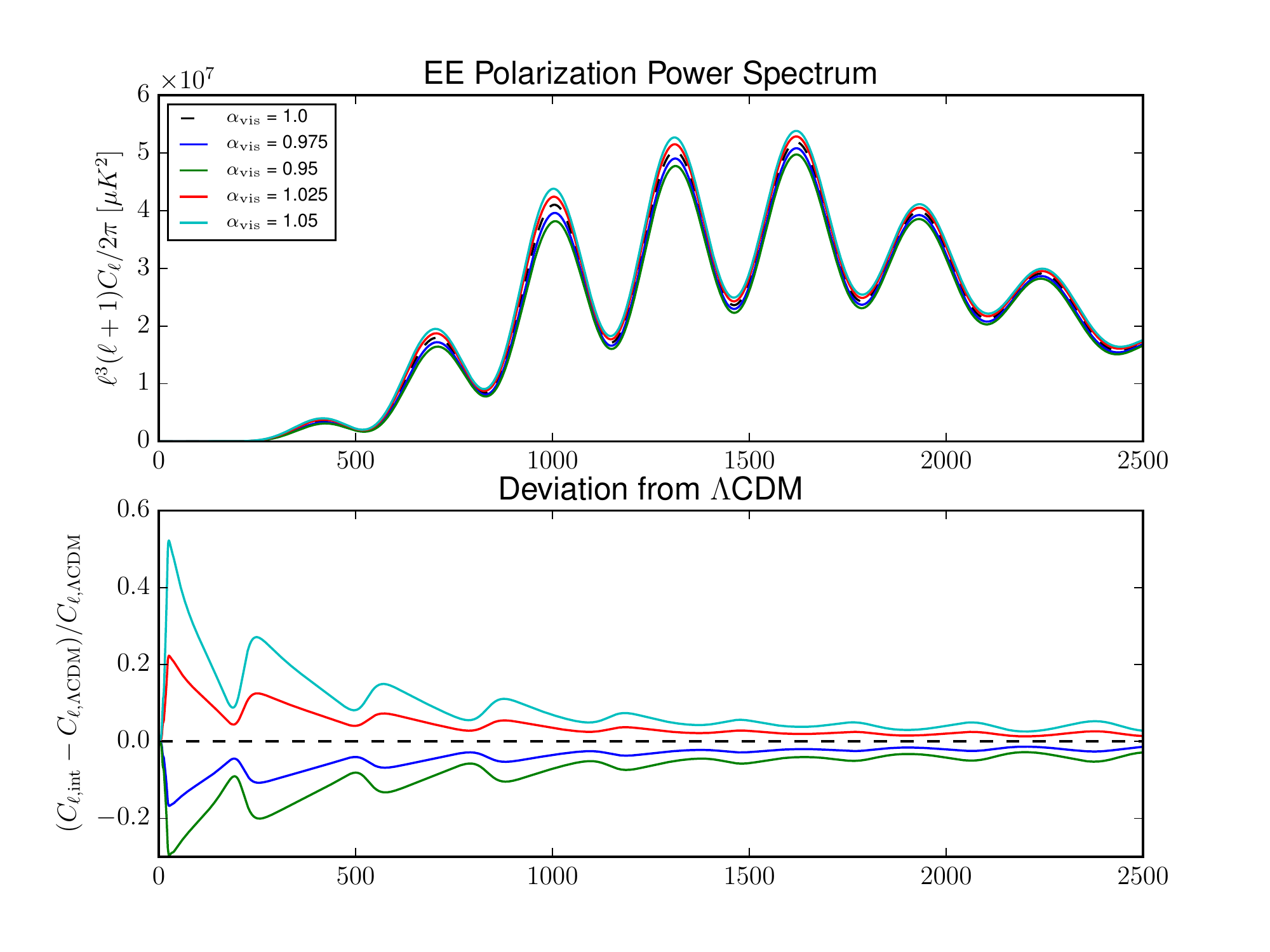}
      \caption{Effects on the polarization power spectrum for different widths of 
	the last scattering surface.}
      \label{fig:der_alpha}
\end{figure*}

In Fig. \ref{fig:der_alpha}, we can see the effects of varying the width of the
last scattering surface on the EE polarization power spectrum. 
As discussed in \cite{harari}, we observe that
 the amplitude of the polarization power spectrum increases roughly quadratically
 with
$\alpha_{\rm vis}$, while the peak locations are almost unchanged. We further see that
 the effect
is stronger on the large angular scales, while the smaller scales are affected
by the so called damping tail, which leads to a suppression of the amplitudes. 
 
We have also studied the effect of varying the width of the visibility function on
the temperature power spectrum and have found that, as expected, the resulting
deviation is much smaller. On scales $\ell > 200$, the fractional difference is less
than 10\% when we perturb the width by $\Delta \alpha_{\rm vis} =0.05$.
On larger scales, $\ell < 200$,  contrary to the case of polarization, the effect on the
power spectrum is quite negligible.

\subsection{First Set: Constraints from Silk Damping and Low $\ell$ 
Polarization-to-Temperature Ratio}
As a first step, we test how well the current measurements of the
temperature and large-scale polarization power spectra are able to constrain the 
duration of last scattering
through $\alpha_{\rm vis}$. Measurements of the temperature power spectrum
constrain the width because as argued in \cite{harari}, the 
longer the photon last scattering lasts for, the stronger the damping effects would 
be on the
smaller scales. On the other hand, the low $\ell$ likelihood 
gives us the polarization-to-temperature ratio on large scales, and thus
also helps us measure the width.

In Table \ref{tab:fits}, we present the best-fit values for the parameters
in the $\Lambda {\rm CDM}$ model and our new parameter $\alpha_{\rm vis}$.
Notice that despite the fact that the low $\ell$ data are affected by cosmic variance 
and systematics,
the large-scale polarization data along with the measurements of the temperature 
power spectrum from the $Planck$
team on small scales, sensitive to
second-order effects such as the damping,  can constrain the width of the
last scattering surface to very good precision (about 0.6\%).

We can translate the measurement of the error of 
$\alpha_{\rm vis} = 0.9937_{-0.0062}^{+0.0056}$
into a measurement of the precision to which we can constrain the
 width of the last scattering surface. To do so, we multiply
it by the value we obtained for its width from our theoretical prediction
($\Delta \eta_\ast = 19 \ {\rm Mpc}$). We find that our current measurements can
constrain the value of $\Delta \eta_\ast$ to within $\sigma [\Delta \eta_\ast] \approx
 \textbf{0.11 {\rm Mpc}}$ (68\% CL),
which is indeed very precise. The slightly lower value of $\alpha_{\rm vis}$ is most likely 
due to
a combination of noise and cosmic variance, but it
might also be suggesting
that in our patch of the Universe, the polarization-to-temperature ratio (effectively) 
happens
to be smaller than the overall, assuming the Standard Model is correct. 

\begin{center}
\begin{table}[htbp!]
\centering
\scalebox{1.1}{
    \begin{tabular}{l | c | c | c | c}
    \hline
     Param & best fit & mean$\pm\sigma$ & 95\% low & 95\% up \\ 
    \hline \hline
    $\alpha_{\rm vis}$ &$0.993$ & $0.994 \pm 0.006$ & $0.981$ & $1.005$ \\
    $\Omega_{b } h^2$ &$0.0222$ & $0.0222 \pm 0.00025$ & $0.02171$ & $0.02268$ \\
    $\Omega_{cdm } h^2$ &$0.118$ & $0.118 \pm 0.002$ & $0.114$ & $0.122$ \\
    $100*\theta_{s }$ &$1.042$ & $1.042 \pm 0.0004$ & $1.041$ & $1.043$ \\
    $ln10^{10}A_{s }$ &$3.08$ & $3.08 \pm 0.03$ & $3.03$ & $3.13$ \\
    $n_{s }$ &$0.972$ & $0.969 \pm 0.006$ & $0.957$ & $0.981$ \\
    $\tau_{reio }$ &$0.0697$ & $0.0682 \pm 0.015$ & $0.04$ & $0.0965$ \\
    $H_0$ &$68.2$ & $68.0 \pm 0.9$ & $66.2$ & $69.9$ \\
    \hline
    \end{tabular}} \\
$-\ln{\cal L}_\mathrm{min} =5355.23$, minimum $\chi^2=1.071e+04$ \\
\caption[Table title text]{Constraints from Silk damping and low $\ell$ 
polarization-to-temperature ratio.}
\label{tab:fits}
\end{table}
\end{center}

\subsection{Second Set: Constraints from Polarization-to-Temperature Ratio on 
all Scales}
Equation \ref{eq:rat} shows that the polarization-to-temperature
ratio depends strongly on the width of the last scattering surface, i.e. 
$C_\ell^{EE} / C_\ell^{TT}
\propto \Delta \eta_\ast^2 \propto \alpha_{\rm vis}^2$. For this reason, after including 
the measurements on the
polarization of the power spectrum, we arrive at even more stringent 
constraints on the value of $\alpha_{\rm vis}$.
These are shown in Table \ref{tab:fits2}. The standard deviation of
$\alpha_{\rm vis}$ is nearly two times smaller
compared with the first set, which corresponds to a precision of
$\sim 0.3\%$). Translating
the value of $\alpha_{\rm vis} = 0.9974_{-0.0034}^{+0.0034}$ into 
the more physical quantity $\Delta \eta_\ast$, we find that the
width of the visibility function is constrained to within $\sigma [\Delta \eta_\ast] 
\approx \textbf{0.065 {\rm Mpc}}$ at 68\% CL.

\begin{center}
\begin{table}[htbp!]
\centering
\scalebox{1.1}{
    \begin{tabular}{l | c | c | c | c}
    \hline
    Param & best fit & mean$\pm\sigma$ & 95\% low & 95\% up \\ \hline \hline
    $\alpha_{\rm vis}$ &$0.996$ & $0.997 \pm 0.0034$ & $0.991$ & $1.004$ \\
    $\Omega_{b } h^2$ &$0.0222$ & $0.0223 \pm 0.0002$ & $0.0219$ & $0.0226$ \\
    $\Omega_{cdm }h^2$ &$0.120$ & $0.119 \pm 0.0015$ & $0.116$ & $0.122$ \\
    $100*\theta_{s }$ &$1.042$ & $1.042 \pm 0.0003$ & $1.041$ & $1.043$ \\
    $ln10^{10}A_{s }$ &$3.03$ & $3.07 \pm 0.03$ & $3.02$ & $3.12$ \\
    $n_{s }$ &$0.960$ & $0.965 \pm 0.005$ & $0.956$ & $0.975$ \\
    $\tau_{reio }$ &$0.0453$ & $0.0666 \pm 0.013$ & $0.0422$ & $0.090$ \\
    $H_0$ &$67.2$ & $67.7 \pm 0.65$ & $66.4$ & $69.0$ \\
    \hline
    \end{tabular}} \\
$-\ln{\cal L}_\mathrm{min} =5355.23$, minimum $\chi^2=1.071e+04$ \\
\caption[Table title text]{Constraints from polarization-to-temperature ratio on all scales.}
\label{tab:fits2}
\end{table}
\end{center}

\begin{figure}
[H]
      \includegraphics[width=3.5in]{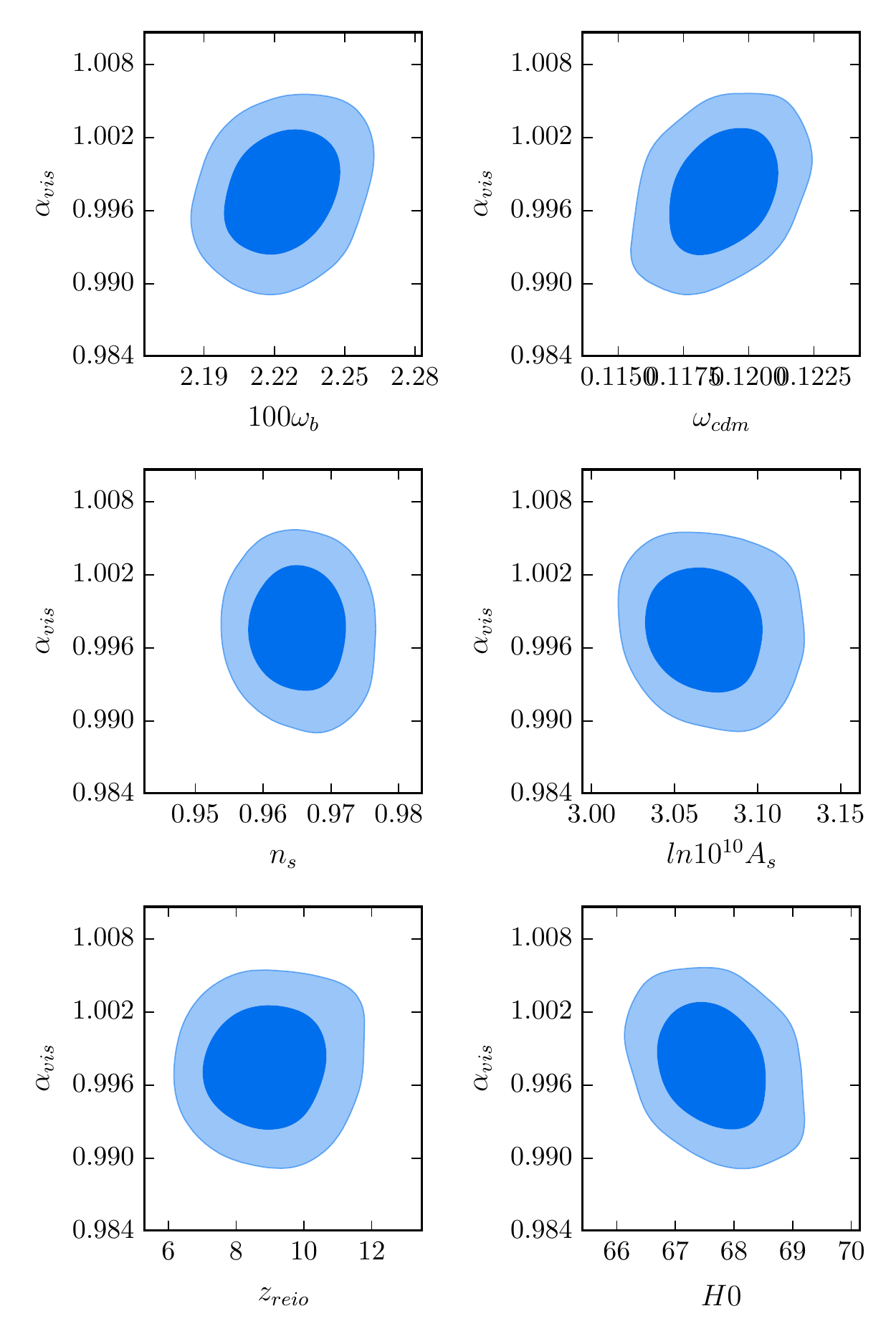}
      \caption{2D contours of $\alpha_{\rm vis}$ and the 6 standard parameters. 
The new parameter seems weakly correlated with
      the other parameters.}
      \label{fig:alpha6}
\end{figure}

In Fig. \ref{fig:alpha6}, we show the correlations between $\alpha_{\rm vis}$
and the 6 standard parameters. Notice that while there appears to be
a weak correlation between $\alpha_{\rm vis}$ and the dark matter density 
$\omega_{\rm cdm}$
and $\alpha_{\rm vis}$ and the Hubble parameter, the dependence overall is 
not very strong, i.e.
the changing of $\alpha_{\rm vis}$ produces a different effect
on the power spectrum than any of the other 6 parameters. 

\section{Correlation with Extra Parameters}
We further tested the conjecture
that $\alpha_{\rm vis}$ measures a new physical effect on the CMB power spectrum by
including parameters beyond the six standard ones in our Monte-Carlo sampling 
engine and looking
for degeneracies. The parameters we added were the effective number of neutrino
species $N_{\rm eff}$ ($\equiv N_{ur}$) and the energy density of curvature 
$\Omega_k$. 

\begin{figure}[H]
      \includegraphics[width=3.5in]{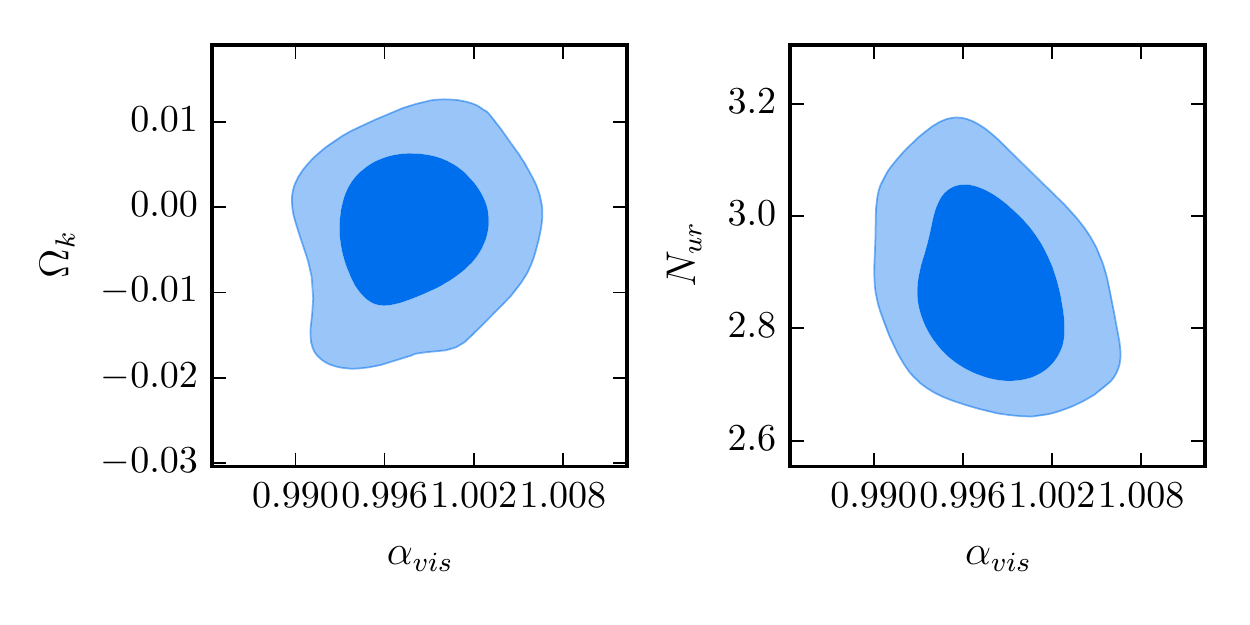}
      \caption{2D contours of $\alpha_{\rm vis}$ and the two extra parameters: the effective
        number of ultra-relativistc species, $N_{ur}$, and the energy density of curvature
        $\Omega_k$. The new parameter does not seem to exhibit strong correlation with
	either parameter.}
      \label{fig:alpha2}
\end{figure}

In Fig. \ref{fig:alpha2}, we show the 2D contours
of $\alpha_{\rm vis}$ and the two extra parameters. 
We do not find degeneracies with
any of the parameters, which supports the claim that $\alpha_{\rm vis}$ provides 
an independent test of the Standard Model
as well as a test of the systematics of a given data set. We find that the mean 
value of $\alpha_{\rm vis}$
is within 1$\sigma$ of the standard prediction for its value 
$\alpha_{\rm vis}=1$.  If future experiments should find that its value differs by 
more than $1 \sigma$ from $\alpha_{\rm vis} = 1$,
that could be indicative of a systematic error or an unaccounted effect on the 
CMB resulting from new physics.

\section{Discussions and Conclusions}
\label{chap:conc}
In this paper, we explored the effect of
 the width of the last scattering surface on the power spectrum and
the constraints we can obtain on its thickness given our current data.
We
 found that with our current measurements of the temperature 
and the large-scale polarization power spectra,
the thickness of the last scattering surface can be constrained to an
astounding precision: $\sigma [\Delta 
\eta_\ast] \approx \textbf{0.11 {\rm Mpc}}$.
The constraint comes from the Silk damping tail,
which gets suppressed as we broaden the width of the
visibility function, and from the large-scale polarization, which is 
strongly dependent on the width.
If we include the polarization data from $Planck$ 2015 on all scales, we get
the much tighter constraint of 
$\sigma [\Delta \eta_\ast] \approx \textbf{0.065 {\rm Mpc}}$.
This is a consequence of the fact that the polarization-to-temperature
ratio is proportional to the width of the last scattering surface squared.
We believe that constraining this parameter
is a good way to test our current model and probe for 
physics beyond the Standard Model.

In the near future, the new polarization data
 from upcoming experiments such as Simons Observatory (SO)
 and CMB-S4 \cite{Abazajian:2016yjj} 
should allow us to measure the parameter $\alpha_{\rm vis}$
with even greater precision. High-$\ell$ data from
SPT and ACT should in principle also help
measure better the diffusion damping tail
and thus put constraints on $\alpha_{\rm vis}$. Including $\alpha_{\rm vis}$ 
when analyzing the new datasets will enable us to
 detect deviations
from the $\Lambda {\rm CDM}$ model
 and look for systematic errors
in these new datasets. In addition, it will be useful when constraining models
which alter the reionization history of the Universe such as self-interacting
 dark matter models and variable-$\alpha$ models.

\begin{acknowledgements}
We are very grateful to Joanna Dunkley for her hard and devoted work
without which the completion of this project would have been
barely possible.
We would like to thank Blake Sherwin for providing us with
useful comments, which helped us refine this paper.
This work was completed as part
of B.~H.'s undergraduate senior thesis.
The Flatiron Institute is supported by the Simons Foundation.
\end{acknowledgements}

\bibliographystyle{apsrev4-1}
\bibliography{alpha}

\end{document}